# Superconductive Ultra-compact Magnetically Coupled Resonator with Twin-spiral Structure

A. S. Averkin, A. Karpov, A. P. Zhuravel, L.V. Filippenko, V. P. Koshelets, S.M. Anlage, and A.V. Ustinov

*Abstract*—. **We describe a practical design of an ultra-compact on-chip superconductive micro-resonator as a potential magnetic element of metamaterial. The achieved resonator size with respect to the wavelength in our experiment is about $\lambda/14400$. The resonator consists of two superconducting Nb spirals, sandwiched face to face, with a small gap filled with dielectric. The spirals are turning in opposite directions: one clockwise, another counter-clockwise. We study the resonator spectral response and its inner modes using numerical simulation in HFSS. In experiment with a Laser Scanning Microscope (LSM) we confirm the predictions for the resonance frequencies of Nb superconductive resonator and its inner modes structures. Small size and the ease of manufacturing make the two-spiral resonator an attractive solution for superconductive filters, coupling circuits, and as a magnetic component of a metamaterial.**

## I. Introduction

A NUMBER of exciting applications of a medium with simultaneous negative permittivity $\varepsilon$ and permeability $\mu$ was theoretically described by V.G. Veselago in 1968 [1]. About 30 years later, a practical example of a material with negative effective permeability $\mu$ was demonstrated by Pendry et al. [2]. In order to create the negative permeability $\mu$, Pendry et al. used an array of split ring resonators (SRR). SRRs interacts mainly with the magnetic component of the electromagnetic (EM) field and gives the possibility to create a medium with effective negative $\mu$. The first demonstration of a medium with both negative $\varepsilon$ and $\mu$ was done by Smith et al. in 2000 [3]. They called this artificial media a metamaterial. The metamaterial was constructed of layers of copper SRRs and layers of wires. SRRs have strong coupling with the magnetic component of the EM field, while the wires have a strong coupling with the electric component of the EM field, acting as electrical dipoles.

The spiral resonator, as compared to an SRR, has equally strong coupling to magnetic field, but much smaller size compared to the resonant wavelength than an SRR, because of the dense placement of turns. Earlier experiments [4], [5] were made with planar spirals made of thick Cu films (~0.35 mm thick in [4] and ~0.25 mm thick in [5]) on dielectric substrates. Such a thick coating is required to minimize the Ohmic losses, making the design intermediate between 2-D and 3-D. Another approach to reduce the size of the spiral resonator is the introduction of the structure with two sandwiched spirals, studied recently by Chen et al. [6]. The demonstrated twin-spiral magnetic metamaterial component size is below $\lambda/1300$ [6].

A further miniaturization of normal-metal spiral resonators has its natural limitation due to the scaling of Ohmic dissipation with spiral width and thickness [7], [8]. In order to demonstrate a deep sub-wavelength size resonator, it appears promising to utilize ultra-compact superconductive spiral resonators. A superconducting spiral resonator and one-dimensional planar array of spiral resonators made of Nb film was studied by Kurter et al. [9]. This resonator has a fundamental resonance frequency of 74 MHz, and its size is as small as $\lambda/675$.

In this work we achieve a strong reduction of the resonator size, using the two superconductive Nb spirals, sandwiched face to face with a small gap in between. The two spirals of the resonator have a strong capacitive coupling. As a result, the frequency of the fundamental mode is lowered by more than 20 times with respect to a single resonator. The resonator is produced by photolithography. We demonstrate that the resonator diameter may take only a small fraction of the wavelength size, below $1/14400$.

## II. Two-spiral Micro-resonator Design

We consider the two planar Archimedean spirals identical in shape, but turning in opposite directions (clockwise and counter-clockwise), and superposed with a thin dielectric layer in between (Fig. 1). The two ring-shaped Archimedean spirals (with no central part) are well coupled to the external RF

This work was supported by the Government of Russian Federation through Goszadanie research project grant 3.2007.2014/K.

One of us (A.K.) acknowledges support of the Government of Russian Federation through grant "Organisation of research work" in Goszadanie.

A.S.A. acknowledges financial support of the Ministry of Education and Science of the Russian Federation in the framework of Increase Competitiveness Program of MISiS.

A.P.Z. acknowledges support from the NASU under Grant program on Nanostructures, Materials, and Technologies.

Corresponding author: Alexandre Karpov (alexandre.karpov@yahoo.com)

Alexander S. Averkin, Alexandre Karpov are with National University of Science and Technology (MISIS), Leninskiy prosp. 4, 119049 Moscow, Russia.

Alexander P. Zhuravel is with B. Verkin Institute for Low Temp. Physics and Engineering, NAS of Ukraine, 61103 Kharkov, Ukraine.

Valery P. Koshelets, Lyudmila V. Filippenko are with Kotel'nikov Institute of Radio Engineering and Electronics, Moscow 125009, Russia.

Steven M. Anlage is with CNAM, Physics Department, University of Maryland, College Park, Maryland 20742-4111, USA

Alexey V. Ustinov is with Physikalisches Institute, Karlsruhe Institute of Technology (KIT), 76131 Karlsruhe, Germany, with National University of Science and Technology (MISIS), Leninskiy prosp. 4, 119049 Moscow, Russia, and Russian Quantum Center (RQC), 100 Novaya St., Skolkovo, Moscow region, 143025, Russia.

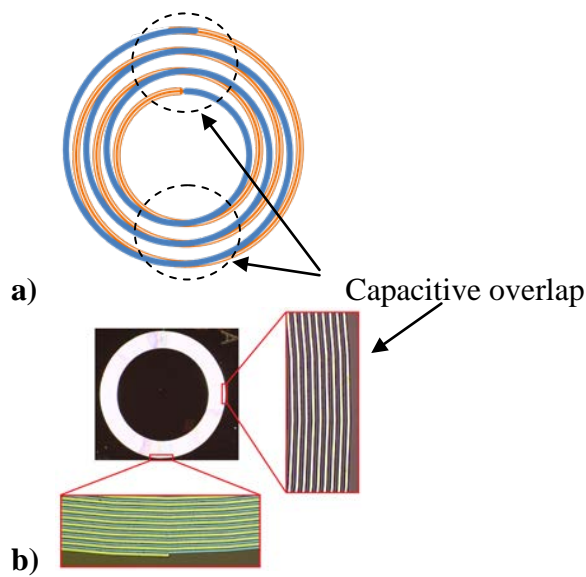

**a)**

Capacitive overlap

**b)**

Fig. 1. (a) A sketch of the two-spiral resonator. The two planar Archimedean spirals are superposed with a thin dielectric layer in between. Spirals are identical in shape and turning in opposite directions: one clockwise, another counter-clockwise. Both spirals are shaped as a ring, with no central part. The spiral arms overlaps play a role of interlayer capacitors, when the other sections of the spiral are still highly inductive, as in a free standing spiral. The additional capacitive loading leads to the reduction of speed of propagation of the signals along the spirals. The spirals act as a distributed RF resonator with multiple modes, and the reduction of the phase speed in the structure gives a lower resonance frequency. The two-spiral structure is made by photo lithography as a three layer circuit. (b) A photo of the two-spiral resonator used in our experiment. At the left blow-up image the windings of the two spirals are matched (overlapped), forming capacitive elements, at the bottom blow-up image the windings are mismatched because the two spirals are turning in the opposite directions.

magnetic field and act as distributed resonant structures. At the visible intersections of the spiral lines there are areas with overlap of the spirals with a strong capacitive coupling along the spiral line. The two spirals have a strong RF coupling and have common resonant modes. The two-spiral on-chip resonator is made as a three layer printed circuit in a single Nb process at Si substrate. The two Archimedean spirals have outer diameter 3 mm and with inner diameter 2.2 mm, 40 turns of Nb wire 5 um wide. The spirals are separated with 300 nm SiO2 layer and turning in opposite directions (one clockwise, another counter-clockwise).The resulting resonator remains a distributed structure, with multiple modes resonances, where the first resonance frequency is much lower than the frequency of a single spiral resonator. The RF experiments with a two-spiral superconductive resonator are presented in the next section.

### III. EXPERIMENT

In experiment we measure the inner modes resonance frequencies and the RF current distributions for the 4 first inner modes of the resonator. The RF test setup for Nb superconducting resonator is cooled in a cryostat to the temperature of about 4.2 K. The RF characterization of the two-spiral resonator is made in a circuit similar to Fig. 2. There the resonator is magnetically coupled to the two RF loop probes. The coupling of the two probing loops (S21) depends on the distance between them and is set relatively small, at about -60 - -80 dB. The superconductive resonator is placed in between the magnetic probes. At the resonance frequency of the superconducting spirals the coupling between the two probes strongly increases [9, 10, 11].

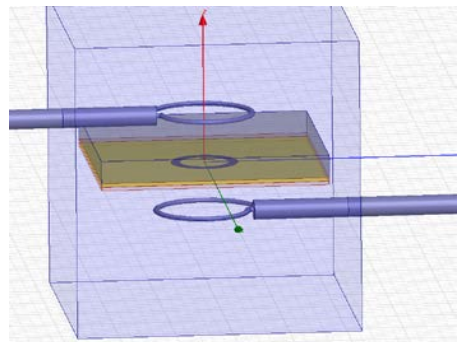

Fig. 2. Experimental RF test setup as simulated numerically in HFSS. A sample of twin-spiral resonator is placed between the two RF loop probes. The two loops probes are weakly magnetically coupled with each other and with the ring shaped resonator. At the resonance frequencies of the two-spiral structure the loop coupling (S21) increases substantially (Fig. 3, 5).

The measured response of the two-spiral resonator is presented in Fig. 3. The multiple sharp resonance peaks are well visible in the data. The first resonance frequency of the twin-spiral resonator is about 6.9 MHz, about 12 times lower than in the same size single spiral resonator studied previously [11]. The opposite winding still allows the resonator coupling with RF magnetic field perpendicular to the resonator plane. As result, one can see strong coupling of the odd modes ($f_1$=6.9 MHz, $f_3$=51 MHz and $f_5$=120 MHz) with coupling increase by 20-30 dB. The even modes ($f_2$=26 MHz, $f_4$=82 MHz), where the RF field is confined mostly to the resonator plane, are weaker coupled to the setup probes. The developed deep subwavelength resonator may be used as meta-atom in RF metamaterial experiments. It is interesting to note, that at the first mode resonator diameter is only $\lambda/14400$ fraction of the wavelength, making the meta-atom compactness comparable to one in real atoms. As a matter of reference we could take the Bohr radius of a Hydrogen atom (0.051 nm) and the first line in Balmer series (653.3 nm) resulting in $\lambda/D$ of about 6100.

The shape of the standing waves of RF current in the superconducting resonator was probed with a cryogenic laser scanning microscope (LSM) [12]. The method is based on the measurement of transmission of RF signal through a thin film superconductive circuit at cryogenic temperature. In order to create an image of the RF currents, the superconductivity of the circuit material is locally depressed by a laser beam, inducing a decrease in the measured RF transmission. The laser beam induced variation of transmission is not uniform, and is greater in the areas with higher RF current amplitude. The mapping of the variation in transmission induced by laser beam is providing the distribution of the RF currents in the circuit.

An example of the laser microscope arrangement is shown

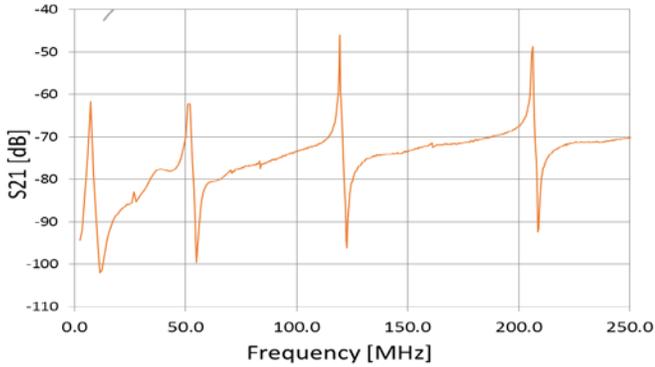

Fig. 3. Measured transmission |S21| of twin-spiral resonators at T=4.5 K in the test setup configuration shown in Fig. 2. The two Archimedean spirals have outer diameter 3 mm and with inner diameter 2.2 mm, 40 turns of Nb wire 5 um wide. The spirals are separated with 300 nm $SiO_2$ layer and turning in opposite directions (one clockwise, another counter-clockwise). The opposite winding allows the resonator coupling with RF magnetic field perpendicular to the resonator plane. As result, the coupling of the odd modes ($f_1$=7 MHz, f3=51 MHz and $f_5$=120 MHz) is stronger, compared to the even modes ($f_2$=26 MHz, f4=82 MHz). The first resonance frequency of the two-spiral resonator ($f_1$=7 MHz) is more than 20 lower, compared to the single spiral resonator of the same size [11].

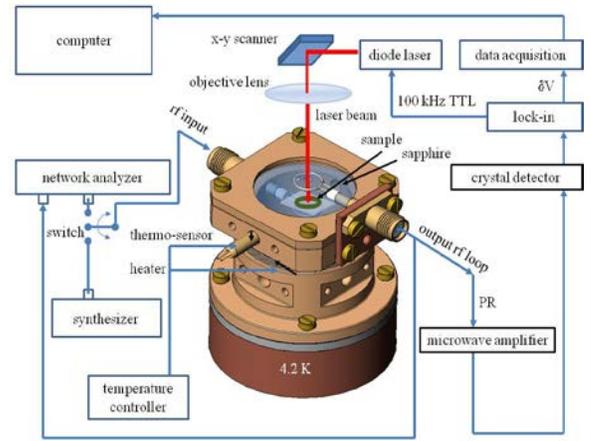

Fig. 4. Schematic of the laser microscope test setup arrangement. A spiral resonator is mounted in the cryostat between the two magnetic loop probes. The probing signal is injected through the upper loop and received at the lower magnetic probe. The sample is scanned with a focused laser beam, generating locally the quasi-particles and inducing the loss. The induced loss is also proportional to the local amplitude of RF current. When mapping the laser beam induced losses over the sample (with RF signal applied), we obtain

in Fig. 4 [12]. The sample is placed in the holder at the surface of a sapphire disk in between of the two RF magnetic loop probes. The sample holder is installed in vacuum inside of a cryostat with optically transparent window. The measurements are performed at temperature of about 4.2 K, and the temperature is controlled using thermo-sensor and heater.

In order to identify the resonance frequencies of the two-spiral resonator, where the LSM measurement should be done, the RF transmission in setup is measured with Agilent E5071C network analyzer coupled to the probes. When the frequency is identified, the setup RF input switch is turned to a computer controlled Anritsu MG37022A RF signal synthesizer, and the setup output port is connected via a Mini Circuits ZX60-3018G+ microwave amplifier to Agilent 8471D crystal diode detector used for detection of the LSM photo-response signal.

The LSM measured RF current images of the first 4 inner modes of the two-spiral resonator are plotted in upper part of the Fig. 5. The measured resonance frequencies are $f_1$=6.9 MHz, $f_2$=26 MHz, $f_3$=51 MHz, $f_4$=82 MHz, and $f_5$=120 MHz. The measured RF current distribution is center-symmetrical, where nodes (bright spots) and antinodes (dark) are located at circular lines. In the area of capacitive overlap the nodal structure is fading. The number of the antinodes corresponds to the order number of the resonance observed. A bright background outside of the resonator is due to a strong photo-response in the Si substrate of our sample. One may note that the two-spiral resonant structure is behaving as a distributed resonator. An intriguing question of the limits of the first mode resonance frequency manipulation in two-spiral resonator in respect to a single spiral case is studied using the numerical simulation in the next section.

## IV. HFSS SIMULATION

We calculated the RF current distribution for the several modes of the designed resonator (bottom of the Fig. 5) and the dependence of the fundamental resonance frequency versus the thickness $d$ of the dielectric-filled gap between the spirals (Fig.6).

In order to verify the possibility of reduction of the fundamental resonance frequency in two-spiral resonator compared to a single spiral and to validate the experimental data, we use the ANSYS High Frequency Structural Simulator (HFSS) [13]. The Driven Mode HFSS program is used for calculating the resonance frequencies and the shape of the standing waves of the resonant modes for spirals used in our experiments. In HFSS calculations the planar spirals are assumed to be made of infinitely thin, lossless metal layer and to be situated at the substrate with dielectric permittivity $\varepsilon_r$. The HFSS simulated circuit structure follows the experimental setup of Fig 2, with the twin spiral resonator sample placed in between of the two weakly coupled magnetic loops. In order to ensure a weak coupling of the two loops, the inter-loop distance is set at two loop radii. In HFSS model the two loop coupling $S_{21}(f)$ is calculated as a function of frequency $f$, and the two-spiral resonator resonance frequencies $f_n$ are determined as the spikes in the $S_{21}(f)$ data.

First, we analyzed the inner modes of the tested two-spiral resonator with the dimensions, as listed in Fig. 3. The HFSS calculated resonance frequencies of the studied two-spiral resonator are in a reasonable agreement with experimental data for the 1st-4th resonant modes (Fig. 5).
The simulated RF current distribution for the first inner modes of the resonator is plotted at the bottom of Fig. 5. The images of RF currents are obtained as the plots of the amplitude of tangent component of magnetic field in the plane just over the surface of the resonator. Note that the simulated current distributions have azimuthal symmetry unlike the LSM data, affected by non-uniformity of filling of the substrate over inductive and the capacitive overlap areas. The number of antinodes is equal to the mode number. In general, the HFSS model of the distributed inner modes in the two-spiral

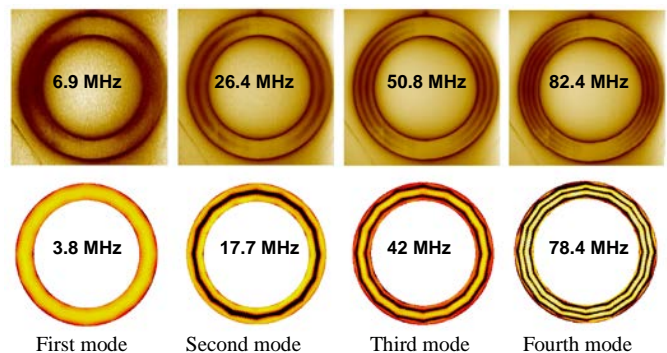

Fig. 5. RF current distribution measured with LSM (above) and HFSS simulated (below) for the first 4 inner modes of the two-spiral resonator. The brighter color corresponds to the higher RF current density. The number of antinodes is equal to the mode number. The capacitive overlap areas do not have pronounced nodal structure in experimental data. The bright background in experimental data is due to a strong photo response in the Si substrate of the tested sample. The measured and calculated resonance frequencies are listed in the middle of each plot.

resonator is confirmed in LSM experiment.

The distance $d$ between the two spirals in the resonator is very important parameter, rapidly changing the first resonance frequency of the twin-spiral resonator (Fig. 6). When the distance between the two spirals is of the order or below of the width of the spiral line $w$, the capacitive overlap areas play an important role (see Fig. 1), scaling the resonance frequency as the square root of the distance. In this case, the capacitance affects the phase speed in the distributed transmission line formed by the two spirals. When the distance $d$ exceeds the width of the lines ($d$ above 15 µm), the capacitance of the overlap is less important, and the resonance frequency goes as the logarithm of the distance, as the inductance of a bifilar line formed by the parallel sections of the spirals. The measured resonance of the first mode $f_1$=6.9 MHz is in a reasonable agreement with the HFSS simulation data.

V. SUMMARY

An ultra-compact superconducting micro-resonator is proposed, HFSS simulated, produced and tested. The achieved wavelength to resonator size ratio is about 14400, making the meta-atom compactness comparable to one in real atoms. The resonator consists of two identical superconducting Nb spirals, sandwiched face to face, with a small gap. The opposite winding of spirals allows the resonator coupling with RF magnetic field perpendicular to the resonator plane. The resonator inner modes structure was properly predicted in numerical model and verified experimentally with the scanning laser microscope. Small size and the ease of manufacturing make the twin-spiral resonator an attractive solution for superconductive filters, coupling circuits, on-chip antennas, and as magnetic component of a metamaterial.

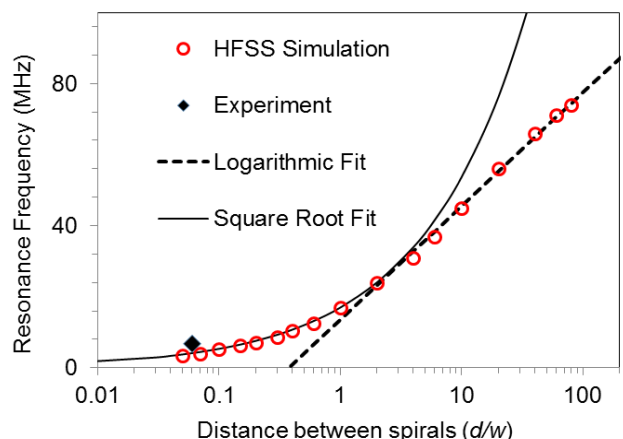

Fig. 6. Twin-spiral resonator fundamental resonance frequency as a function of the ratio of distance ($d$) between the two spirals to the width of the spiral line (w) simulated with HFSS (open circles). All other twin-spiral dimensions are the same, as presented in Fig. 3. For the spacing larger than 15 µm ($d/w$ = 3) the resonance frequency dependence versus $d$ is logarithmic, and below the 10 µm distance ($d/w$ = 2) the resonance frequency goes as a square root of the distance. The measured resonance frequency of 6.9 MHz (Fig. 3) fits reasonably well the HFSS simulated data for the distance $d$ of 300 nm, typical for the thickness of dielectric layer in the tested sample. One can note the 12 fold reduction of the fundamental resonance frequency in the two spiral resonator compared to the single spiral resonance frequency of about 83 MHz.